\documentclass[aps,prd,twocolumn,superscriptaddress,preprintnumbers,showpacs,amsmath,amssymb]{revtex4}
\usepackage{epsf}
\usepackage{latexsym}
\usepackage{amsmath}
\usepackage{amsfonts}
\usepackage{amssymb}
\usepackage{graphicx}
\newcommand{\be}{\begin{eqnarray}}
\newcommand{\ee}{\end{eqnarray}}
\newcommand{\ud}{\mathrm{d}}

\newcommand{\lp}{\ell_{\rm p}}
\newcommand{\mpl}{M_{\rm p}}
\newcommand{\mh}{M_{\rm H}}
\newcommand{\md}{M_{(5)}}
\newcommand{\ld}{\ell_{(5)}}
\newcommand{\mew}{M_{\rm ew}}
\newcommand{\meff}{M_{\rm eff}}
\newcommand{\Meff}{M_{\rm eff}}
\newcommand{\geff}{g_{\rm eff}}
\newcommand{\mc}{M_{\rm c}}
\newcommand{\rc}{r_{\rm c}}
\newcommand{\alphac}{\alpha_{\rm c}}
\newcommand{\mmax}{M_{\rm max}}
\newcommand{\rem}{R_{\rm EM}}
\newcommand{\reff}{R_{\rm eff}}
\newcommand{\rh}{R_{\rm H}}

\begin{document}
\preprint{Preprint number: DO-TH-09/19}
\title{Theoretical survey of tidal-charged black holes at the LHC}
\author{Roberto~Casadio}
\email{casadio@bo.infn.it }
\affiliation{Dipartimento di Fisica, Universit\`a di
Bologna and I.N.F.N., via Irnerio 46, 40126 Bologna, Italy}
\author{Sergio~Fabi}
\email{fabi001@bama.ua.edu}
\affiliation{Department of Physics and Astronomy, The University
of Alabama, Box 870324, Tuscaloosa, AL 35487-0324, USA}
\author{Benjamin~Harms}
\email{bharms@bama.ua.edu}
\affiliation{Department of Physics and Astronomy, The University
of Alabama, Box 870324, Tuscaloosa, AL 35487-0324, USA}
\author{Octavian~Micu}
\email{octavian.micu@tu-dortmund.de}
\affiliation{Fakult\"at f\"ur Physik, Technische Universit\"at Dortmund,
D-44221 Dortmund, Germany}
\begin{abstract}
We analyse a family of brane-world black holes which solve the effective four-dimensional
Einstein equations for a wide range of parameters related to the unknown bulk/brane physics.
We first constrain the parameters using known experimental bounds and, for the allowed
cases, perform a numerical analysis of their time evolution, which includes accretion through
the Earth.
The study is aimed at predicting the typical behavior one can expect if such black holes
were produced at the LHC.
Most notably, we find that, under no circumstances, would the black holes reach
the (hazardous) regime of Bondi accretion.
Nonetheless, the possibility remains that black holes live long enough to escape
from the accelerator (and even from the Earth's gravitational field) and result in missing
energy from the detectors.
\end{abstract}
\pacs{04.70.Dy, 04.50.+h, 14.80.-j}
\maketitle
%
\section{Introduction}
\label{intro}
The existence of extra spatial dimensions~\cite{arkani,RS} and a sufficiently small
fundamental scale of gravity opens up the possibility that microscopic
black holes can be produced and detected~\cite{bhlhc,CH,BHreview} at the
Large Hadron Collider (LHC).
Since the existence of large extra dimensions permits the formation of
microscopic black holes, these large extra dimensions and black holes will be
searched for at the LHC.
Therefore it is important to study all of the implications of the Randall-Sundrum (RS)
model for black hole production and decay at the LHC.
In this paper we shall, in particular, consider the RS brane-world of Ref.~\cite{RS}.
Our world is thus a three-brane (with coordinates $x^\mu$, $\mu=0,\ldots,3$)
embedded in a five-dimensional bulk with the metric
\be
\ud s^2 =e^{-|y|/\ell}\,g_{\mu\nu}\,\ud x^{\mu}\,\ud x^{\nu} + \ud y^2
\ ,\label{RS}
\ee
where  $y$ parameterizes the fifth dimension and $\ell$ is a length determined by the
brane tension.
This parameter relates the four-dimensional Planck mass $\mpl$ to the five-dimensional
gravitational mass $\md$ and one can have $\md\simeq 1\,$TeV$/c^2$
(for bounds on $\ell$, see, e.g., Ref.~\cite{harko}) and black holes with mass in the
TeV~range.
Note that, experimental limits require $\md\gtrsim 1\,$TeV but there is no strong
theoretical evidence that places $\md$ at any specific value below $\mpl$.
The brane must also have a thickness, which we denote by $L$,
below which deviations from the four-dimensional Newton law occur.
Current precision experiments require that $L \lesssim 44\,\mu$m~\cite{Lbounds},
whereas theoretical reasons imply that
$L\gtrsim\ld\simeq \lp\,\mpl/\md\simeq 2\cdot 10^{-19}\,$m.
In the analysis below, the parameters $\md$ and $L$ are assumed to be
independent of one another, but within the stated ranges.
\par
Despite many efforts, to date, only approximate black hole metrics are known on the
brane~\cite{dadhich,bwbh,BHreview}.
In a previous publication~\cite{CH}, we showed that, using a specific form of
the metric found in Ref.~\cite{dadhich}, and a specific choice of parameter
values, black hole lifetimes can be very long.
It was then conjectured~\cite{plaga} that such black holes might be able to grow
to catastrophic size within the Earth, contrary to the picture~\cite{giddings}
that arises in the ADD scenario~\cite{arkani}.
This possibility was then refuted in Ref.~\cite{GM2} and,  in Ref.~\cite{bhEarth1},
we solved the system of equations which describes the mass of a black hole and
its momentum as functions of time for various initial conditions and values of the
critical mass which occur in that model.
\par
In the present paper we consider a wider class of metrics of the form
obtained in Ref.~\cite{dadhich} and constrain the parameters which appear in
it in order to use one form for a wider range of black hole masses.
The constraints will follow from the experimental bounds mentioned at the
beginning and will us allow to restrict the space of parameters to a
manageable range.
Within this range, we will study the evolution of the corresponding black holes
numerically and several conclusions will be obtained.
Most remarkably, we shall see that tidal-charged black holes produced at
the LHC would very likely evaporate instantaneously and, even for those values
of the parameters which lead to an initial growth, no catastrophic scenario
will arise.
Life-times could however be long enough to allow for black holes to
escape from the detectors and result in significant amounts of missing
energy.
This would be a very strong signature of micro-black holes at the LHC.
\par
We shall use units with $1=c=\hbar=\mpl\,\lp=\ld\,\md$,
where $\mpl\simeq 2.2\cdot 10^{-8}\,$kg and $\lp\simeq 1.6\cdot 10^{-35}\,$m
are the Planck mass and length related to the
four-dimensional Newton constant $G_{\rm N}=\lp/\mpl$.
In our analysis we shall consider only the five-dimensional RS~scenario
with $\md\simeq \mew\simeq 1\,$TeV ($\simeq 1.8\cdot 10^{-24}\,$kg),
the electro-weak scale, corresponding to a fundamental length
$\ld\simeq 2.0\cdot 10^{-19}$m.
\section{Brane-world black hole metrics}
\label{metrics}
Since gravity propagates in the bulk, a matter source located on the brane will
give rise to a modified energy momentum tensor in the Einstein equations projected
on the three-brane~\cite{shiromizu}.
By solving the latter, one finds that this backreaction can be described in the
form of a tidal ``charge'' $q$, and the effective four-dimensional metric for
a brane-world black hole should thus be given by~\cite{dadhich}
\be
\ud s^2 =
- A\,\ud t^2 + A^{-1}\,\ud r^2 + r^2\left(\ud\theta^2 +\sin(\theta)^2\,\ud\phi^2\right)
\ ,
\label{tidal}
\ee
with
\be
A=1-\frac{2\,\lp\,M}{\mpl\,r}-q\,\frac{\mpl^2\,\lp^2}{\md^2\,r^2}
\ .
\ee
For $q>0$, this metric has one horizon at
\be
\rh=\lp\left(\frac{M}{\mpl}+\sqrt{\frac{M^2}{\mpl^2}+q\,\frac{\mpl^2}{\md^2}}\right)
\ .
\label{rhq}
\ee
It is then plausible that both the Arnowitt-Deser-Misner (ADM) mass $M$
and the (dimensionless) tidal charge $q$ depend
upon the black hole proper mass $M_0$ in such a way that when
$M_0$ vanishes, so do $M$ and $q$.
The functions $M=M(M_0)$ and $q=q(M_0)$ could only be determined precisely
by solving the full bulk equations, for example using the four-dimensional
metric~\eqref{tidal} as a boundary condition.
Unfortunately, this task cannot be performed exactly, but only
numerically or perturbatively~\cite{bwbh,mazza}.
\subsection{Parametrized tidal charge}
In order to simplify the analysis, we shall first assume that $M=M_0$ and,
at least for $M\sim \md$, that the functional form of $q$ is given by~\footnote{A
survey of the possible bulk features corresponding to different relations $q=q(M)$
is currently being performed using the method of Ref.~\cite{mazza}.}
\be
q \simeq \left(\frac{\mpl}{\md}\right)^\alpha\left(\frac{M}{\md}\right)^\beta
\ ,
\ee
where $\alpha$ and $\beta> 0$ are real parameters.
In general, the tidal charge $q$ can also depend on the thickness of the brane
$L$ via some dimensionless function of $L$ and possibly the four-dimensional
and five-dimensional values of the Planck length $\lp$ respectively $\ld$.
The analysis of this case is deferred to a separate study, which is a work
in progress.
The tidal charge dependence on the brane tension, and on the bulk cosmological
constant are already encoded in the dependence of $q$ on the five-dimensional
Planck mass.
The length scale $\ell$ in Eq.~\eqref{RS} is determined by the brane tension
via junction conditions~\cite{RS},
and relates the four-dimensional fundamental constants to the
five-dimensional ones.
Returning to our case of study,
it is possible that the parameters $\alpha$ and $\beta$ depend on
the mass scale as well, for example if there occurs a dimensional phase
transition of the form described in Ref.~\cite{bhEarth1}
(see also Section~\ref{a0b1} below).
In the present paper, we shall mostly assume that no such case occurs and
constrain $\alpha$ and $\beta$ by using
$\md\simeq \mew \simeq 1\,$TeV$/c^2$ and the known bounds on $L$.
For this purpose, we note that the tidal term in the metric,
\be
A_{\rm t}
\simeq
\left(\frac{\mpl}{\md}\right)^{\alpha+2}\left(\frac{M}{\md}\right)^\beta
\frac{\lp^2}{r^2}
\ ,
\label{At}
\ee
dominates over the usual General Relativistic term,
\be
A_{\rm N}\simeq 2\,\frac{M\,\lp}{\mpl\,r}
\ ,
\label{AN}
\ee
for $r\lesssim \rc$, with
\be
\rc\simeq
\lp\left(\frac{\mpl}{\md}\right)^{3+\alpha}
\left(\frac{M}{\md}\right)^{\beta-1}
\ .
\label{rc}
\ee
It then makes sense to require that $\rc$ be shorter than the length scale
above which corrections to the Newton potential have not yet been detected.
That is, we impose
\be
\rc\ll L
\ ,
\label{rcc}
\ee
if the black hole is ``small'', in the sense that
\be
\rh\ll\rc\ll L
\ .
\label{sbh}
\ee
If the black hole were ``large'', meaning that
$\rc\ll\rh$, the constraint~\eqref{rcc} could actually be evaded,
but this case is of no interest here.
In fact, for $\rh\ll\rc$, the horizon radius can be estimated using the
tidal contribution~\eqref{At},
\be
\rh\simeq
\lp\left(\frac{\mpl}{\md}\right)^{1+\alpha/2}\left(\frac{M}{\md}\right)^{\beta/2}
\ ,
\label{smallRh}
\ee
otherwise $\rh$ approaches the usual four-dimensional expression
\be
\rh\simeq
2\,\lp\,\frac{M}{\mpl}
\ .
\label{largeRh}
\ee
Note then that the condition of classicality for the horizon, namely
\be
\rh\gg\lambda_M
\ ,
\ee
where
\be
\lambda_M\simeq\ld\,\frac{\md}{M}=\lp\,\frac{\mpl}{M}
\ee
is the Compton length of the black hole,
for small black holes with $\alpha\simeq 0$ reads
\be
M\gg \md
\ ,
\ee
whereas for large black holes the condition approaches the usual
four-dimensional condition
\be
M\gg \mpl
\ .
\ee
This implies that micro-black holes with a mass in the TeV range
will always be small in the above sense that $\rh\ll\rc$.
\par
A more refined classicality condition for all values of $\beta$ and
$\alpha$ can actually be obtained from
the effective four-dimensional Euclidean action~\cite{CH,gergely},
\be
S_{(4)}^{\rm E}
=
\frac{\mpl\,(4\,\pi\,\rh^2)}{16\,\pi\,\lp}
\ .
\ee
For small black holes, the above expression can be approximated
by
\be
S_{(4)}^{\rm E}\simeq \frac{\lp\,\mpl}{4}\left(\frac{\mpl}{\md}\right)^{\alpha+2}
\left(\frac{M}{\md}\right)^\beta
\ ,
\label{sS}
\ee
which can be rewritten as
\be
S_{(4)}^E=\lp\,\mpl\,\tilde M^{\beta}
\ ,
\ee
with $\tilde M=M/\Meff$ and
\be
\Meff=
{\md}\left[\frac{1}{4}
\left(\frac{\mpl}{\md}\right)^{\alpha+2}\right]^{-\frac{1}{\beta}}
\ .
\ee
The area law then implies that the degeneracy of a black hole is
counted in units of $\Meff$ (see Eq.~\eqref{areaL} and
Refs.~\cite{mfd,bc2}).
A black hole is classical if its mass is much larger than $\meff$,
which implies that $\meff$ must be no larger than
$\md$ in order to have TeV~scale black holes.
Using the fact that $\beta$ is positive, the above relation allows us to
impose a lower bound on $\alpha$ for all values of $\beta$, namely
\be
\alpha\gtrsim -2
\ .
\label{alpha-2}
\ee
Also note that, for large black holes the Euclidean action given in
Eq.~\eqref{sS} will approach the usual four-dimensional expression
\be
S_{(4)}^{\rm E}\simeq \lp\,\frac{M^2}{\mpl}
\ .
\label{lS}
\ee
\par
Regarding Eq.~\eqref{sS}, it is interesting to note that its functional
dependence on $M$ is the same as that of a $(4+d)$-dimensional
Schwarzschild black hole~\cite{CH,giddings} with
\be
\beta=\frac{d+2}{d+1}
\ ,
\ee
so that, from the thermodynamics point of view, small tidal-charged
black holes with $1<\beta<2$ mimic higher-dimensional Schwarzschild
black holes~\footnote{This qualitative observation will be later
supported by studying the time evolution in Section~\ref{evol}.}.
\subsection{Parameter space for small black holes}
We now proceed to analyze different ranges of $\beta>0$.
Since we are interested in micro-black holes, we shall consider
only ``small'' holes which satisfy the condition~\eqref{sbh}.
For $\beta\not=1$, the condition $\rc=L$ corresponds to a critical mass
\be
\mc=
\md\left[\frac{L}{\lp}\left(\frac{\md}{\mpl}\right)^{3+\alpha}
\right]^{\frac{1}{\beta-1}}
\ .
\label{mc}
\ee
The case $\beta=1$ will be analysed separately in Section~\ref{beta1}.
Further, for $\beta\not=2$, the condition that $\rc=\rh$ leads to
\be
M\simeq
\mh
\equiv
\md\left(\frac{\md}{\mpl}\right)^{\frac{\alpha+4}{\beta-2}}
\ ,
\label{mh}
\ee
whereas for $\beta=2$ one finds no constraint on $M$ but
$\rh\ll\rc$ implies that
\be
\left(\frac{\md}{\mpl}\right)^{\alpha+4}\ll 1
\ ,
\ee
or $\alpha>-4$.
The case with $\beta=2$ will also be analyzed in detail in
Section~\ref{beta=2}.
\par
Let us now look at the conditions in Eqs.~\eqref{rcc} and \eqref{sbh}
for $\beta\neq 1$ and $\beta\neq 2$, so that $\mc$ and $\mh$ are
properly defined as above.
The condition~\eqref{rcc} for the critical radius to be smaller than the thickness
of the brane implies
\be
\left(\frac{M}{\md}\right)^{\beta-1}
\ll
\frac{L}{\lp}\left(\frac{\md}{\mpl}\right)^{3+\alpha}
\ .
\label{rc_cond}
\ee
We then have two cases:
(i)~for $0<\beta<1$, the above condition yields
\be
M\gtrsim\mc
\ ,
\label{M>Mc}
\ee
while, (ii)~for $\beta>1$,
\be
M\lesssim\mc
\ ,
\label{M<Mc}
\ee
where $\mc$ was given in Eq.~\eqref{mc}.
Similarly, one can analyze the lower bound in Eq.~\eqref{sbh}.
Since we are  interested only in small black holes, we
assume that the condition~\eqref{rc_cond} is satisfied
(the critical radius is smaller than the thickness of the brane).
Below the critical radius the tidal term dominates,
and the Schwarzschild radius of the black hole from Eq.~\eqref{rhq}
can be approximated by the tidal component.
Using this approximation, $\rh\ll\rc$ can be written as
\be
\left(\frac{M}{\md}\right)^{\beta-2}
\gg
\left(\frac{\md}{\mpl}\right)^{\alpha+4}
\ .
\ee
We again have two separate cases:
(a)~for $\beta<2$ we get
\be
M\lesssim\md
\left(\frac{\md}{\mpl}\right)^{\frac{\alpha+4}{\beta-2}}
\equiv\mh
\ ,
\label{betall2}
\ee
and, (b)~for $\beta>2$,
\be
M\gtrsim\md\left(\frac{\md}{\mpl}\right)^{\frac{\alpha+4}{\beta-2}}
\equiv\mh
\ .
\label{betagg2}
\ee
\subsection{ $\beta>1$}
In this range, the tidal term grows with $M$ faster than the Newtonian term and
Eq.~\eqref{rcc} becomes the upper bound~\eqref{M<Mc} on the maximum black
hole mass, namely
\be
M\lesssim\mc
\ ,
\ee
which grows with $L$, as one might have naively expected.
For $M\gtrsim \mc(L)$, one should therefore use a different form for $q$,
which might signal a (dimensional) phase transition of the sort considered,
for instance, in Refs.~\cite{chplb,CH,bhEarth1}.
\par
Note that $\mc\gg\md$ if
\be
\frac{L}{\lp}\gg\left(\frac{\mpl}{\md}\right)^{\alpha+3}
\ ,
\label{LLp}
\ee
or
\be
\alpha\lesssim\alphac
\simeq
\frac{\ln(L/\lp)}{\ln(\mpl/\md)}-3
\ ,
\label{alpha1}
\ee
For $\ld\lesssim L\lesssim 44\,\mu$m and $\md\simeq \mew$, this implies
\be
-2\lesssim \alpha_c\lesssim -1.1
\ ,
\label{alphac}
\ee
so that, for $L\to\ld$, the allowed parameter space becomes empty
[due to the constraint~\eqref{alpha-2}].
We must then analyse the three cases (a), (b) and $\beta=2$ separately.
\subsubsection{ $1<\beta<2$}
In this case, a ``small'' black hole must have a mass in the range
\be
\md\ll M\ll \min\{\mh,~\mc\}
\ ,
\ee
This further implies that both $\mc$ and $\mh$ need to be much larger than
$\md$ for tidal micro-black holes to exist.
The condition for $\mc$ was given in Eq.~\eqref{LLp}, and
requiring that $\mh\gg\md$ means that
\be
\left(\frac{\mpl}{\md}\right)^{\alpha+4}\gg 1
\ ,
\ee
or $\alpha\gtrsim -4$, which is however weaker than~\eqref{alpha-2}.
The condition~\eqref{alpha1} must hold for all values of $\beta>1$,
which means that the allowed range for $\alpha$ would be given by
\be
-2\lesssim \alpha\lesssim \alphac\lesssim -1.1
\ .
\label{ca1}
\ee
\par
However we see that, for $\beta\to 1^+$ and $\alpha$ in the above
range, the critical mass $\mc\to\infty$, and an infinitely massive black hole
would be of the tidal kind, thus ruling out the Schwarzschild geometry.
We therefore require that $\mc\lesssim M_\odot\simeq 10^{54}\,$TeV
(the mass of the sun), which yields
\be
\alpha\gtrsim\alpha_\odot
\simeq
\frac{\ln(L/\lp)-(\beta-1)\,\ln(M_\odot/\md)}{\ln(\mpl/\md)}-3
\ .
\label{alpha_sun}
\ee
This constraint becomes quickly ineffective [that is $\alpha_\odot<-2$
for $\beta\gtrsim 1.3$ and \eqref{ca1} prevails], but will be carefully taken
into consideration in Section~\ref{evol}.
\subsubsection{ $\beta=2$}
\label{beta=2}
In this case the black hole is small and classical if $\alpha\gtrsim -2$
and
\be
M\ll\mc
\simeq
\md\,\frac{L}{\lp}\left(\frac{\md}{\mpl}\right)^{3+\alpha}
\ .
\ee
Then $\mc\gg\md$ again leads to Eq.~\eqref{LLp}.
One therefore concludes that the range of allowed values of $\alpha$
is again given in Eq.~\eqref{ca1}.
Further, as we noted above, the constraint~\eqref{alpha_sun} is already
ineffective.
\subsubsection{ $\beta>2$}
The black hole is ``small'' if
\be
\max\{\mh,~\md\} \ll M\ll \mc
\ .
\ee
The condition that $\mh\ll\mc$ then implies the new constraint
\be
\frac{L}{\lp}
\gg
\left(\frac{\md}{\mpl}\right)^{\frac{\alpha+\beta+2}{\beta-2}}
\ ,
\ee
and, for the usual values for $L$ and $\md$, we obtain
\be
\alpha\gtrsim-3\,\beta+2
\ .
\ee
The stronger bound is again given by the condition~\eqref{alpha-2}
for the black hole to be classical.
Along with the conditions in Eq.~\eqref{alpha1}, the range for
$\alpha$ in this case becomes again that given in Eq.~\eqref{ca1}.
\subsection{ $0<\beta<1$}
In this case, the tidal term grows with $M$ more slowly than the Newton
potential and we obtained
\be
M\gtrsim \mc
\ ,
\ee
which, correspondingly, decreases for increasing $L$.
We assume that the black hole is created with a mass close to the
five-dimensional Planck mass $\md$.
This implies that for the black hole to be tidal, the critical mass
$\mc$ needs to be smaller than $\md$, which results in the upper
bound~\eqref{alpha1}.
\par
The black hole is then small for
\be
\md\ll M\ll \mh
\ ,
\ee
where $\mh$ is again given in Eq.~\eqref{mh}.
A necessary condition again is that $\mh\gg\md$ if $\alpha\gtrsim -4$.
Combining all the restrictions, we again arrive at the range in Eq.~\eqref{ca1}.
\begin{figure}[t]
\centering
\epsfxsize=3.2in
\epsfbox{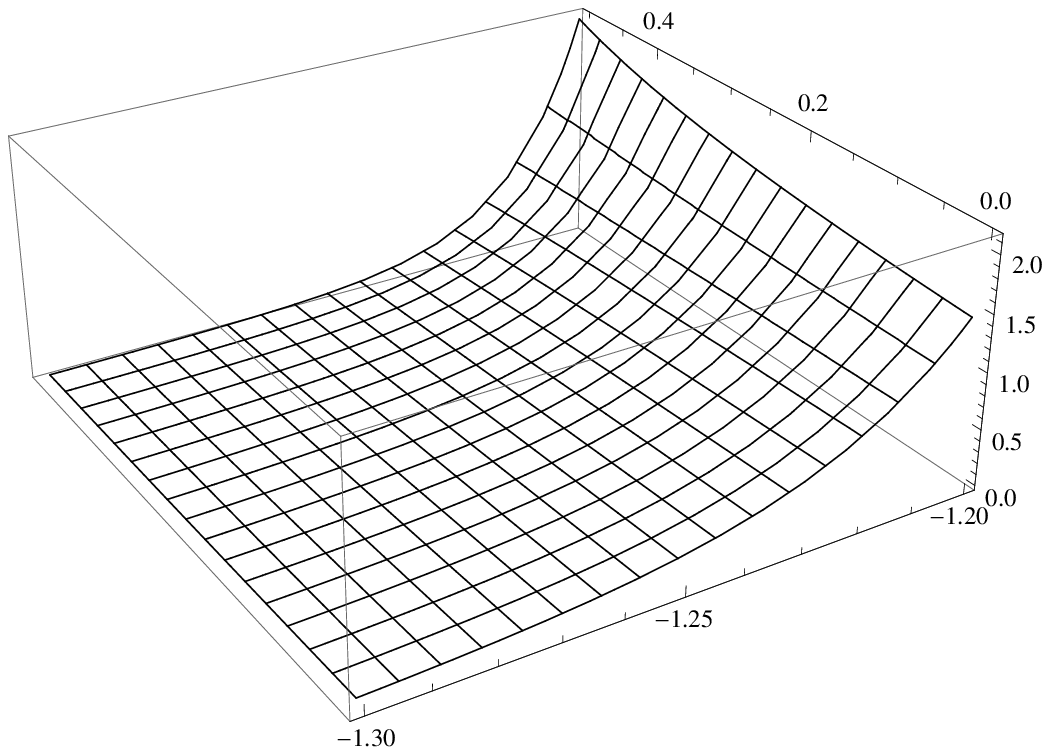}
\caption{Remnant mass $\mc$ in TeV$/c^2$ for $\md\simeq \mew$,
$L\simeq 1\,\mu$m,
$-1.3<\alpha<-1.2$ and $0<\beta<0.5$.}
\label{Mcm}
\end{figure}
\subsection{ $\beta=1$}
\label{beta1}
Both terms in the metric coefficient $A$ now grow linearly with $M$ and
Eq.~\eqref{rc} reads
\be
\rc\simeq
\lp\left(\frac{\mpl}{\md}\right)^{3+\alpha}
\ ,
\label{rc1}
\ee
which does not depend on $M$ and, therefore, Eq.~\eqref{rcc} does not
place any bound on $M$.
It can instead be used to constrain the parameter $\alpha$, namely
Eq.~\eqref{alpha1}.
\subsubsection{ $\alpha=\alphac$}
It is interesting to work this case in detail.
For $\alpha=\alphac$ one obtains
\be
A=1-\frac{2\,\lp\,M}{\mpl\,r}-\frac{\lp\,L\,M}{\mpl\,r^2}
\ ,
\ee
and
\be
\rh=
\lp\left(\frac{M}{\mpl}+\sqrt{\frac{M^2}{\mpl^2}+\frac{L\,M}{\lp\,\mpl}}\right)
\ .
\ee
For the usual choice of $M\sim M_{(5)}\simeq \mew\simeq 1\,$TeV and
$L\lesssim 44\,\mu$m,
the second term in the square root above dominates and the horizon radius
is well approximated by
\be
\rh \simeq \lp\,\sqrt{\frac{L\,M}{\lp\,\mpl}}
\ ,
\label{R5}
\ee
which, for $M\gtrsim M_{(5)}\simeq \mew$ and $L\gg \lp$, is larger than the
four-dimensional Schwarzschild radius~\eqref{largeRh}.
\subsubsection{ $\alpha=0$}
\label{a0b1}
This case was employed in Refs.~\cite{CH,plaga,bhEarth1} and we note here
that it corresponds to
\be
\rc
\simeq
\lp\left(\frac{\mpl}{\md}\right)^{3}
\ ,
\ee
which, for $\md\simeq\mew$, is much larger than all the allowed values of $L$.
Consequently, this case can only be used for sufficiently small mass $M$ such
that the gravitational force of the black hole is negligible small.
This condition can be realised by requiring that the capture radius of
the black hole on the surrounding matter is much smaller than $L$,
which yields $\mc\lesssim 1\,$kg~\cite{bhEarth1}.
As was explained in Ref.~\cite{bhEarth1}, when $M$ approaches $\mc$, one
expects a ``dimensional phase transition'', which can be rephrased by
saying that the functional dependence of the tidal charge $q$ on $M$ must
change.
We shall not consider this case here any further, and just refer the
reader to Ref.~\cite{bhEarth1} for more details.
\section{Evaporation}
\label{evaporation}
We shall describe black hole evaporation by means of the microcanonical
ensemble~\cite{mfd,bhEarth1}.
The general form for the luminosity of a black hole in $D$ space-time
dimensions is given by
\be
{\cal L}_{(D)}(M)
=
\int_0^\infty \sum_{s=1}^S n_{(D)}(\omega)\,
\Gamma_{(D)}^{(s)}(\omega)\omega^{D-1}d\omega
\ ,
\label{LH}
\ee
where $D$ is the space-time dimensionality,
$\Gamma_{(D)}^{(s)}$ the grey body factor with $S$ the number of particle
species which can be emitted.
For the sake of simplicity, $\sum_{s}\Gamma_{(D)}^{(s)}$ will be taken
to be a constant.
\par
The occupation number density for the Hawking particles in the microcanonical
ensemble is in general given by~\cite{mfd,bc2}
\be
n_{(D)}=B\sum_{n=1}^{[[M/\omega]]}\,
\exp
\left\{\frac{S_{(D)}^{\rm E}(M-n\,\omega)}{\lp\,\mpl}
-\frac{S_{(D)}^{\rm E}(M)}{\lp\,\mpl}\right\}
,
\label{n}
\ee
where $S_{(D)}^{\rm E}$ is the Euclidean action,
$[[X]]$ denotes the integer part of $X$ and $B=B(\omega)$
encodes deviations from the area law~\cite{r1} (in the following we
shall also assume $B$ is constant in the range of interesting values of $M$).
As we noted before, since
\be
\frac{S_{(D)}^{\rm E}(M)}{\lp\,\mpl}
=
\left(\frac{M}{\Meff}\right)^\beta
\equiv
\tilde M^\beta
\ ,
\label{areaL}
\ee
the black hole degeneracy is counted in units of $\Meff$.
For the usual Schwarzschild action~\eqref{lS}, $n(\omega)$ mimics the canonical
ensemble (Planckian) number density in the limit $M \to \infty$, and the luminosity
becomes
\be
{\cal L}_{\rm H}
\sim
\int_0^\infty\frac{\omega^{3}\,\ud\omega}{e^{\beta_{\rm H}\,\omega}\mp 1}
\sim
T_{\rm H}^{4}
\ ,
\ee
where $T_{\rm H}=\beta_{\rm H}^{-1}=1/(8\,\pi\,M)$ is the Hawking temperature.
Upon multiplying by the horizon area [see Eq.~\eqref{largeRh}], one then obtains
the Hawking evaporation rate~\cite{hawking}
\be
\frac{\ud M}{\ud\tau}
\simeq
\frac{\geff\,\mpl^3}{960\,\pi\,\lp\,M^2}
\ ,
\label{eva4}
\ee
where $\geff\simeq 10$ is the typical number of effective degrees of freedom
into which a four-dimensional black hole can evaporate.
\par
\begin{figure}[t]
\centering
\raisebox{1.6in}{$s$}
\epsfxsize=3.0in
\epsfbox{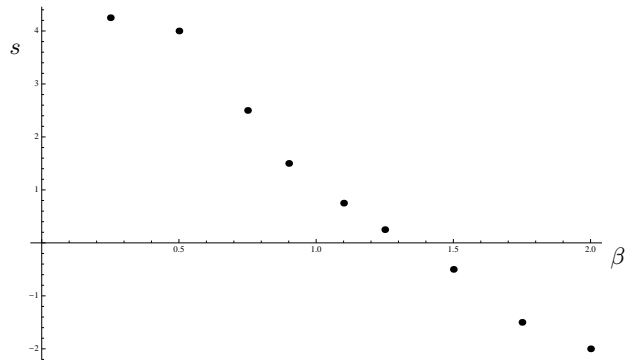}
\raisebox{0.5in}{$\beta$}
\caption{Power $s$ in Eq.~\eqref{evax} for $0<\beta\le 2$.}
\label{plotMx}
\end{figure}

We now calculate the luminosity for the general case of the tidal-charged
black holes.
Since we are working in an effective four-dimensional picture, we can set
$D=4$ in the above expressions.
From Eqs.~\eqref{n} and \eqref{LH}, we get
\be
{\cal L}(M)
=
B\,e^{-\tilde M^\beta}\,
\int_0^\infty\sum_{n=1}^{[[\tilde M/\tilde\omega]]}
e^{\left({\tilde M-n\,\tilde\omega}\right)^\beta}
\tilde\omega^3\,\ud\tilde\omega
\ ,
\ee
with $\tilde M$ defined earlier, $\tilde\omega=\omega/\Meff$ and
all numerical constants (independent of $M$) included in $B$.
Upon further redefining $B$ at each step, one then finds
\be
\!\!\!\!\!\!
{\cal L}(M)
&\!\!=\!\!&
B\,e^{-\tilde M^\beta}\,
\sum_{n=1}^{\infty}\int_0^{\tilde M/n}
e^{\left({\tilde M-n\,\tilde\omega}\right)^\beta}
\tilde\omega^3\,\ud\tilde\omega
\nonumber
\\
&\!\!=\!\!&
B\,e^{-\tilde M^\beta}\,
\sum_{n=1}^{\infty}\frac{1}{n^4}
\int_0^{\tilde M}
e^{x^\beta}\left(\tilde M-x\right)^3 \ud x
 .
\label{LHbeta}
\ee
The integral in Eq.~\eqref{LHbeta} can be evaluated analytically for fixed
$\beta$, but explicit expressions can be rather cumbersome and will be
omitted in general.
Since we require the classicality condition $M\gg\Meff$, the decay rate
is in general well-approximated by a power-law, namely
\be
\left.\frac{\ud M}{\ud\tau}\right|_{\rm evap}
\simeq
C\,M^s
\ ,
\label{evax}
\ee
where a sample of the powers $s$ is plotted in Fig.~\ref{plotMx}.
The normalization in the above expression will be fixed by the same
procedure as in Refs.~\cite{CH,bc2}.
We shall therefore equate the rate~\eqref{evax} with the Hawking
expression~\eqref{eva4} at the mass scale $M=\mc$ in Eqs.~\eqref{M<Mc}
and \eqref{M>Mc} above (or below) which brane-world corrections are
negligibly small.
\par
For instance, let us work out the case with $\beta=1$ and
$\alpha=\alphac$ in detail.
The effective four-dimensional Euclidean action~\cite{CH,gergely} is given by
\be
S_{(4)}^{\rm E}
=
\frac{\mpl\,(4\,\pi\,\rh^2)}{16\,\pi\,\lp}
\simeq
\frac{L\,M}{4}
=
\lp\,\mpl\left(\frac{M}{\meff}\right)
\ ,
\label{SE}
\ee
with
\be
\meff =4\,\mpl\,\frac{\lp}{L}
\ ,
\ee
and, given the limits on $L$ we discussed in the Introduction, we have
a rather wide range for $\meff$, namely
\be
10^{-14}\,{\rm TeV}\lesssim \meff\lesssim\mpl
\ .
\ee
Finally, note that both $\alphac$ and $\md$ have been replaced by the
phenomenological length $L$ in all of the relevant expressions.
The luminosity in this case is simple enough, that is
\be
{\cal L}
\simeq
B\,e^{-\tilde M}\,
\sum_{n=1}^{\infty}\frac{1}{n^4}
\int_0^{\tilde M}
e^{x}\left(\tilde M-x\right)^3 \ud x
\simeq
\tilde B
\ ,
\label{Lq}
\ee
where we used $\meff\ll\mew\sim M$~\footnote{This approximation fails
when $M\sim\meff\simeq 1\,$eV.
However, the actual decay of a black hole is a discrete process which
causes jumps in $M$ and, for such low masses, it is clear that continuous
equations are not a reliable approximation.}
and $\tilde B$ is a new constant.
Upon multiplying by the horizon area [see Eq.~\eqref{R5}], we then get
the microcanonical evaporation rate per unit proper time
\be
\left.\frac{\ud M}{\ud\tau}\right|_{\rm evap}
\simeq
C\,M
\ ,
\label{eva5}
\ee
where $C$ is again a constant.
We then equate the rate~\eqref{eva5} with the Hawking expression~\eqref{eva4}
for $M=\mc$ defined by $\rh(\mc)\simeq L$.
Eq.~\eqref{R5} then yields
\be
\mc\simeq
\mpl\,\frac{L}{\lp}
\ .
\ee
Finally
\be
C=
\frac{\geff}{960\,\pi\,\lp}\left(\frac{\mpl}{\mc}\right)^3
\simeq
\frac{\geff\,\lp^2}{960\,\pi\,L^3}
\ ,
\label{C1c}
\ee
where we used Eq.~\eqref{mc} in the approximate equality.
\section{Subatomic accretion}
\label{accretion}
There are two basic mechanisms by which a microscopic black hole in general
might accrete:
one due to the \emph{collisions} with the atomic and sub-atomic particles encountered
as they sweep through matter, and one due to the gravitational force the black
hole exerts on surrounding matter once it comes to rest.
The latter form is known as \emph{Bondi accretion} and is appreciable only when the
black holes have horizon radii greater than atomic size.
\par
In our analysis we focus on the subatomic mechanism, whose fundamental
equation is given by
\be
\left.\frac{\ud M}{\ud t}\right|_{\rm acc}
=
\pi\,v\,\rho\,\reff^2
\ ,
\label{acc}
\ee
where $\rho$ is the density of the material through which the black hole is moving,
and $v$ is the relative velocity of the black hole and the surrounding matter,
while $t$ is the time of observers at rest with respect to the medium.
\par
For sufficiently small horizon radius $\rh$, the capture radius $\reff$ can
be determined by simple Newtonian arguments.
In particular, we can assume that it is given by the range over which the
gravitational force of the black hole can overcome the electromagnetic force
which binds the nucleus of an atom to the surrounding medium.
An expression for this electromagnetic capture radius in five dimensions
can be obtained following the analysis of Ref.~\cite{giddings} and using
the metric~\eqref{tidal}.
Upon neglecting the $1/r$ terms, one first obtains a Newtonian force
\be
F_{\rm G}\simeq
-\frac{\lp\,L\,M\,m}{\mpl\,(b-d)^3}
\ ,
\ee
where $m$ is the typical mass of a matter constituent, $b$ the impact parameter
and $d$ the displacement of $m$.
This force must be equated to the electromagnetic restoring force inside the atom,
\be
F_{\rm E}(d)=-K\,d
\ ,
\ee
where $K$ is a constant,
and the resulting equality maximized with respect to $d$ at fixed $b$.
The final result yields the electromagnetic capture radius
\be
\rem
\simeq
\left(\frac{\lp\,L\, M\,m}{K\,\mpl}\right)^{1/4}
=
C_{\rm EM}\,M^{1/4}
\ ,
\label{REM}
\ee
which is meaningful only if $\rem\gg \rh$.
\par
Again for the simple case of $\beta=1$ and $\alpha=\alphac$ this can be expressed as.
\be
M
\ll
\frac{\mpl\,m}{\lp\,L\,K}
\equiv
M_{\rm EM}
\ .
\ee
For example, with $L\simeq 1\,\mu$m, $\md\simeq\mew$, $K=224\,$J$/$m$^2$
and $m\simeq 6\cdot 10^{-27}\,$kg, one obtains
\be
M_{\rm EM}\simeq 10^{22}\,{\rm kg}
\ .
\ee
\par
We can also use the above capture radius to bound the maximum black hole mass
so that deviations from the Newton law at short distance are below the tested scale,
that is $\rem\ll L$.
This yields
\be
M\ll
\frac{K\,L^3\,\mpl}{\lp\,m}
\equiv
M'_{\rm EM}
\ .
\ee
Upon using the same values above, we obtain
\be
M'_{\rm EM}\simeq 10^{21}\,{\rm kg}
\ .
\ee
Since $M'_{\rm EM}\ll M_{\rm EM}$, one can use the capture radius~\eqref{REM} in
the evolution equation~\eqref{acc} up to $M\simeq M'_{\rm EM}$.
\par
In Section~\ref{evol}, we shall see that the above bounds on the mass are actually
irrelevant for our analysis.
\section{Time-evolution}
\label{evol}
The time evolution of the black hole mass is in general
obtained by summing the evaporation and accretion expressions,
\be
\frac{\ud M}{\ud t} =
\left.\frac{\ud M}{\ud t}\right|_{\rm evap}
+\left. \frac{\ud M}{\ud t}\right|_{\rm acc}
\ ,
\label{dMdt}
\ee
where the decay rate in the reference frame of the Earth is related to
the proper decay rate by
\be
\left.\frac{\ud M}{\ud t}\right|_{\rm evap}
\simeq
-\frac{1}{\gamma}
\left.\frac{\ud M}{\ud \tau}\right|_{\rm evap}
\ ,
\label{evapo}
\ee
and $\gamma$ is the relativistic factor for a point-particle of mass $M$
and three-momentum of magnitude $p$,
\be
\gamma =\frac{\sqrt{M^2 + p^2}}{M}
\ .
\ee
Finally the time-evolution of the momentum in the Earth frame is described
by the equation
\be
\frac{\ud p}{\ud t}
=
\frac{p}{M}\left.\frac{\ud M}{\ud t}\right|_{\rm evap}
\ .
\label{dpdt}
\ee
The net change of mass with respect to time~\eqref{dMdt} and the
equation~\eqref{dpdt} for the time evolution of the momentum form
a system of equations which can be solved numerically to obtain $M(t)$ and $p(t)$.
\par
Again, for the simple case of $\beta=1$ and $\alpha=\alphac$,
Eqs.~\eqref{eva5} and \eqref{C1c} yield the evaporation rate
\be
\left.\frac{\ud M}{\ud t}\right|_{\rm evap}
\simeq
-\frac{\geff\,\lp^2\,M^2(t)}{960\, \pi\,L^3\,\sqrt{M^2(t)+p^2(t)}}
\ ,
\ee
and
\be
\frac{\ud p}{\ud t}
\simeq
-\frac{\geff\,\lp^2\,M(t)\,p(t)}{960\, \pi\,L^3\,
\sqrt{M^2(t)+p^2(t)}}
\ .
\ee
The accretion rate is given by Eq.~\eqref{acc} for $\reff=\rem$,
\be
\left.\frac{\ud M}{\ud t}\right|_{\rm acc}
\simeq
\left(\frac{\lp\,L\,m}{K\,\mpl}\,M(t)\right)^{1/2}
\!\!
\frac{\pi\, \rho\,p(t)}{\sqrt{M^2(t)+p^2(t)}}
\ ,
\ee
where $\rho\simeq 5.5\cdot 10^3\,$kg$/$m$^3$ is the Earth's mean density.
Note that accretion dominates only if the momentum is larger than a critical value,
which for this case reads
\be
p_{\rm c}
=\frac{g_{\rm eff}}
{960\,\pi^2\,\rho}
\left(\frac{\lp^{3}\,K\,\mpl\,M}{L^{7}\,m}\right)^{1/2}
\ .
\label{pcr}
\ee
Also note that the evaporation rate grows with $M$ faster than the accretion rate,
which implies that the black hole cannot accrete indefinitely.
\begin{table*}
\centering
\begin{tabular}{|c|c|c|c|c|c|c|c|c|c|c|c|}
\hline
$p(0)$ & $M(0)$  & $\mc$  & $\mmax$ & $\rem$ & $\rh$ & $S$ & $T$
&$M_{\rm E}$ & $R_{\rm E}$ & $t_{\rm E}$ & $v_{\rm E}$
\\
(TeV$/c$) & (TeV$/c^2$) & (TeV$/c^2$) & (kg) & (m) & (m) & (m) & (sec) & (kg) & (m) & (sec) &
(km$/$sec)
\\
\hline
$5.0$&$10.0$& $1\cdot 10^{44}$ &$2.5\cdot 10^{-5}$ & $1.0\cdot 10^{-12}$&$6.0\cdot 10^{-22}$
&$3.8\cdot 10^{15}$ &$2.6\cdot 10^{25}$ &$1.4\cdot 10^{-21}$ &$1.0\cdot 10^{-16}$
&$2.6$ &$1.9\cdot 10^{3}$
\\
\hline
$4.0$ &$10.5$& $1\cdot 10^{44}$ &$1.9\cdot 10^{-5}$ & $1.0\cdot 10^{-12}$
&$5.0\cdot 10^{-22}$ &$2.6\cdot 10^{15}$ &$2.1\cdot 10^{25}$ &$1.4\cdot 10^{-21}$
&$1.0\cdot 10^{-16}$ &$3.3$ &$1.6\cdot 10^{3}$
\\
\hline
$3.0$ &$10.5$& $1\cdot 10^{44}$ &$1.3\cdot 10^{-5}$ & $1.0\cdot 10^{-12}$
&$4.0\cdot 10^{-22}$ &$2.1\cdot 10^{15}$ &$1.5\cdot 10^{25}$ &$1.3\cdot 10^{-21}$
&$1.0\cdot 10^{-16}$ &$4.0$ &$1.3\cdot 10^{3}$
\\
\hline
$2.0$ &$10.8$& $1\cdot 10^{44}$ &$7.6\cdot 10^{-6}$ & $8.0\cdot 10^{-13}$
&$2.8\cdot 10^{-22}$ &$1.6\cdot 10^{15}$ &$1.0\cdot 10^{25}$ &$1.3\cdot 10^{-21}$
&$1.0\cdot 10^{-16}$ &$6.0$ &$8.3\cdot 10^2$
\\
\hline
$1.0$ &$11.0$& $1\cdot 10^{44}$ &$3.0\cdot 10^{-6}$ & $7.0\cdot 10^{-13}$
&$1.6\cdot 10^{-22}$ &$1.0\cdot 10^{15}$ &$5.2\cdot 10^{24}$ &$1.4\cdot 10^{-21}$
&$1.0\cdot 10^{-16}$ &$13$ &$3.9\cdot 10^2$
\\
\hline
$1.0\cdot 10^{-1}$&$11.0$&$1\cdot 10^{44}$ &$1.4\cdot 10^{-7}$ & $3.0\cdot 10^{-13}$
&$2.4\cdot 10^{-23}$ &$2.2\cdot 10^{14}$ &$5.2\cdot 10^{23}$ &$1.4\cdot 10^{-21}$
&$1.0\cdot 10^{-16}$ &$1.3\cdot 10^2$ &$39$
\\
\hline
$1.0\cdot 10^{-2}$ &$11.0$& $1\cdot 10^{44}$ &$6.5\cdot 10^{-9}$ & $1.5\cdot 10^{-13}$
&$3.5\cdot 10^{-24}$ &$4.8\cdot 10^{13}$ &$5.2\cdot 10^{22}$ &$1.4\cdot 10^{-21}$
&$1.0\cdot 10^{-16}$ &$1.3\cdot 10^3$ &$3.9$
\\
\hline
\end{tabular}
\caption{Time evolution of black hole mass as function of initial momentum
for $L=44\,\mu$m, $\beta=1.25$, $\alpha=-1.8$ which result in critical mass
$M_c=10^{44}\,$TeV$/c^2$.
\label{p_0}}
\end{table*}
\par
In the following we shall evolve a black hole produced with
a typical initial mass $M(0)=10\,$TeV$/c^2\simeq 1.8\cdot 10^{-23}\,$kg
and momentum $p(0)\le 5\,$TeV$/c$~\footnote{These
values correspond to a black hole energy in the laboratory of about $11\,$TeV,
which is consistent with the fact that the LHC total collision energy is $14\,$TeV and
the fact that a black hole cannot be the only product of a collision.},
with $K=224\,$J$/$m$^2$ and $m=5.5\cdot 10^{-27}\,$kg.
We will analyze values for the parameter $\beta$ in each of the different ranges
considered in Section~\ref{metrics}.
For the black holes to be ``tidal", limits for the ranges of $\alpha$ will be imposed
according to our findings in that Section, that is
\be
-2\lesssim\alpha\lesssim\alphac\lesssim -1.1
\ ,
\ee
for $\beta\not=1$ or $2$, and \eqref{alpha_sun} will also be implemented for
$1<\beta\lesssim 1.3$.
Besides these parameters which determine the metric, the only free parameter
in the model is given by the size of the extra dimension $L$.
This will be varied in the range
$10^{-2}\,\mu$m$\,\lesssim L\lesssim 44\,\mu$m~\footnote{For shorter values
of $L$, the behavior of tidal-charged black holes approaches the usual
four-dimensional one and is not particularly interesting.},
the corresponding critical mass $\mc$ being given as of Eq.~\eqref{mc}.
Our results are given in Tables~\ref{p_0}-\ref{mcmax0001}, in which
$\mmax$ is the maximum black hole mass, $\rem$ and
$\rh$ the corresponding maximum values of the horizon and capture radius,
$S$ and $T$ the space covered and the time taken to reach $\mmax$;
$M_{\rm E}$ the mass reached after travelling the Earth's diameter,
$R_{\rm E}$ the corresponding capture radius, $t_{\rm E}$ the time
to travel the Earth's diameter and $v_{\rm E}$ the velocity at that point.
\subsection{Rapidly decaying solutions}
\begin{figure}[t]
\centering
\raisebox{1.6in}{$M$}
\epsfxsize=3.0in
\epsfbox{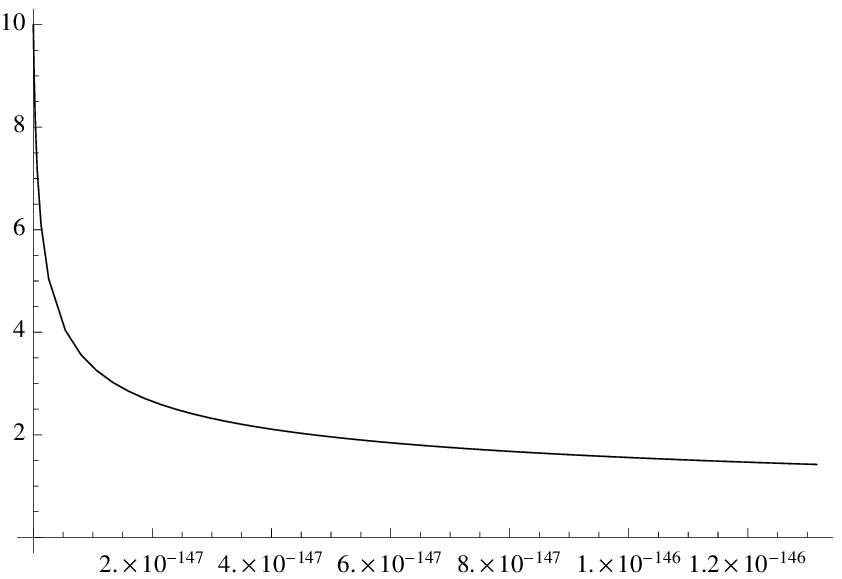}
$t$
\\
\raisebox{1.5in}{$p$}
\epsfxsize=3.0in
\epsfbox{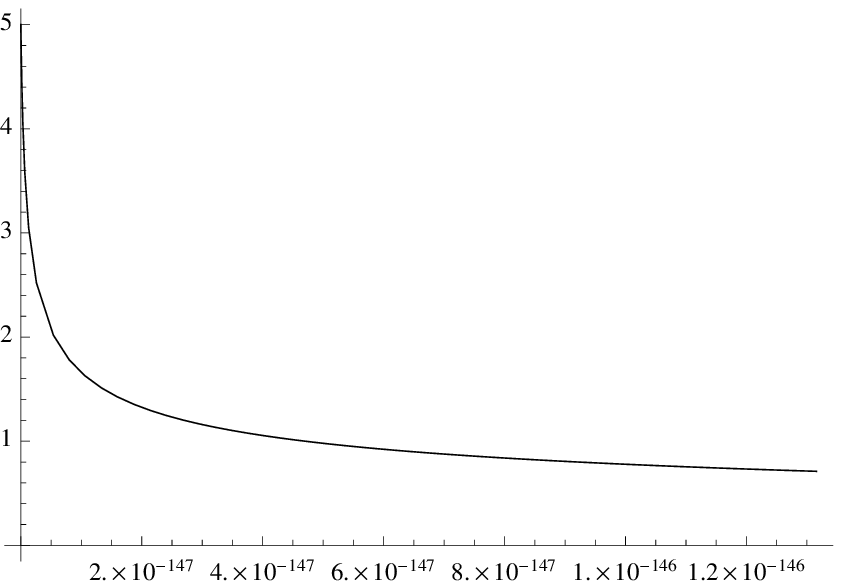}
$t$
\caption{Mass (in TeV$/c^2$) and momentum (in TeV$/c$)
for $L=1\,\mu$m, $\beta=0.5$, $\alpha=-1.5$, $M(0)=10\,$TeV$/c^2$
and $p(0)=5\,$TeV$/c$.}
\label{rapid}
\end{figure}

The first important result is that the \emph{black hole decays instantly}
(i.e., the decay time is shorter than $10^{-10}\,$sec)
\emph{after being created for $0<\beta<1$ and $1.25\lesssim\beta$},
all other parameters being varied within the ranges given previously
in Section~\ref{metrics}~\footnote{The case $\beta=1$ was studied
in Ref.~\cite{bhEarth1} and we refer the reader to that paper for
the details.}.
Fig.~\ref{rapid} shows a typical example of the time evolution of mass
and momentum in this case.
\subsection{Growing solutions}
\begin{table*}[t]
\centering
\begin{tabular}{|c|c|c|c|c|c|c|c|c|c|c|}
\hline
$L$ ($\mu$m) & $\mc$  (TeV$/c^2$)
& $\mmax$  (kg)& $\rem$ (m) & $\rh$ (m)
& $S$ (m) & $T$ (sec)
&$M_{\rm E}$ (kg) & $R_{\rm E}$ (m) & $t_{\rm E}$ (sec) & $v_{\rm E}$
(km/sec)
\\
\hline
5.0& $1\cdot 10^{53}$& $1\cdot 10^{25}$ & $1\cdot 10^{-5}$ &$1\cdot 10^{-4}$
&$1\cdot 10^{31}$ &$1\cdot 10^{70}$ &$8.1\cdot 10^{-22}$ &$4\cdot 10^{-17}$
&$5.2\cdot 10^{-1}$ &$1.1\cdot 10^{4}$
\\
\hline
1.0 & $1\cdot 10^{46}$& $1\cdot 10^{9}$ & $1\cdot 10^{-9}$ &$1\cdot 10^{-13}$
&$1\cdot 10^{23}$ &$1\cdot 10^{47}$ &$8.1\cdot 10^{-23}$ &$1\cdot 10^{-17}$
& $2.3\cdot 10^{-1}$ &$3.2\cdot 10^{4}$
\\
\hline
 $1.0\cdot 10^{-1}$ & $1\cdot 10^{36}$& $1\cdot 10^{-13}$& $1\cdot 10^{-15}$
 &$1\cdot 10^{-25}$ &$1\cdot 10^{12}$ &$1\cdot 10^{14}$ &$3.3\cdot 10^{-23}$
 &$8\cdot 10^{-18}$ & $1.3\cdot 10^{-1}$ &$7.8\cdot 10^{4}$
\\
\hline
 $1.0\cdot 10^{-2}$ & $1\cdot 10^{26}$& N/A&N/A &N/A &N/A &N/A &N/A &N/A &N/A &N/A
\\
\hline
\end{tabular}
\caption{Time evolution of black hole mass as function of extra-dimension size $L$ for
$\beta=1.1$, $\alpha=-1.5$.
Initial conditions are: $M(0)=10\,$TeV$/c^2$ ($=1.8\cdot 10^{-24}\,$kg)
and $p(0)=5\,$TeV$/c$.
N/A means black hole mass does not grow.
\label{L_dependence}}
\end{table*}
\begin{table*}
\centering
\begin{tabular}{|c|c|c|c|c|c|c|c|c|c|c|c|}
\hline
$L$ ($\mu$m)&$\alpha$& $\mc$  (TeV$/c^2$)
& $\mmax$  (kg)& $\rem$ (m) & $\rh$ (m)
& $S$ (m) & $T$ (sec)
&$M_{\rm E}$ (kg) & $R_{\rm E}$ (m) & $t_{\rm E}$ (sec) & $v_{\rm E}$
(km/sec)
\\
\hline
44 &-1.44& $2\cdot 10^{53}$& $1\cdot 10^{25}$ & $1\cdot 10^{-4}$ &$1\cdot 10^{-5}$
&$1\cdot 10^{30}$ &$1\cdot 10^{70}$ &$1.4\cdot 10^{-21}$ &$1\cdot 10^{-16}$
&$2.6$ &$2.0\cdot 10^3$
\\
\hline
5.0& -1.50&$4\cdot 10^{53}$& $1\cdot 10^{25}$ & $1\cdot 10^{-5}$ &$1\cdot 10^{-4}$
&$1\cdot 10^{31}$ &$1\cdot 10^{70}$ &$8.1\cdot 10^{-22}$ &$4\cdot 10^{-17}$
&$5.2\cdot 10^{-1}$ &$1.1\cdot 10^{4}$
\\
\hline
$1.0\cdot 10^{-1}$ & -1.59&$1\cdot 10^{51}$& $1\cdot 10^{19}$& $1\cdot 10^{-7}$
&$1\cdot 10^{-8}$ &$1\cdot 10^{28}$ &$1\cdot 10^{62}$ &$3.3\cdot 10^{-23}$
&$8\cdot 10^{-18}$ &$1.3\cdot 10^{-1}$ &$7.8\cdot 10^{4}$
\\
\hline
$1.0\cdot 10^{-2}$ &-1.67& $1\cdot 10^{54}$& $1\cdot 10^{24}$& $1\cdot 10^{-6}$
&$1\cdot 10^{-6}$ & $1\cdot 10^{32}$&$1\cdot 10^{71}$ &$2.2\cdot 10^{-23}$
&$4\cdot 10^{-18}$ &$1.0\cdot 10^{-1}$ &$1.1\cdot 10^{5}$
\\
\hline
\end{tabular}
\caption{Time evolution of black hole mass with critical mass $\mc$ near upper bound and
$\beta=1.1$.
Initial conditions are: $M(0)=10\,$TeV$/c^2$ ($=1.8\cdot 10^{-24}\,$kg)
and $p(0)=5\,$TeV$/c$.
\label{mcmax}}
\end{table*}
\begin{table*}
\centering
\begin{tabular}{|c|c|c|c|c|c|c|c|c|c|c|}
\hline
$L$ ($\mu$m)& $\mc$  (TeV$/c^2$)
& $\mmax$  (kg)& $\rem$ (m) & $\rh$ (m)
& $S$ (m) & $T$ (sec)
&$M_{\rm E}$ (kg) & $R_{\rm E}$ (m) & $t_{\rm E}$ (sec) & $v_{\rm E}$
(km/sec)
\\
\hline
5.0& $4\cdot 10^{53}$& $1\cdot 10^{23}$ & $1\cdot 10^{-5}$ &$1\cdot 10^{-5}$
&$1\cdot 10^{29}$ &$1\cdot 10^{70}$ &$2.5\cdot 10^{-22}$ &$4\cdot 10^{-17}$
& $3.0\cdot 10^2$ &21
\\
\hline
1.0 & $1\cdot 10^{46}$& $1\cdot 10^{7}$ & $1\cdot 10^{-10}$ &$1\cdot 10^{-14}$
&$1\cdot 10^{22}$ &$1\cdot 10^{47}$ &$1.1\cdot 10^{-22}$ &$1\cdot 10^{-17}$
&$1.6\cdot 10^2$ &51
\\
\hline
$1.0\cdot 10^{-1}$ & $1\cdot 10^{36}$& $1\cdot 10^{-15}$& $1\cdot 10^{-16}$
&$1\cdot 10^{-26}$ &$1\cdot 10^{11}$ &$1\cdot 10^{14}$ &$4.4\cdot 10^{-23}$
&$1\cdot 10^{-17}$ &86 &$1.2\cdot 10^2$
\\
\hline
$1.0\cdot 10^{-2}$ & $1\cdot 10^{26}$& N/A&N/A &N/A &N/A &N/A &N/A &N/A &N/A &N/A
\\
\hline
\end{tabular}
\caption{Time evolution of black hole mass as function of extra-dimension size $L$ for
 $\beta=1.1$, $\alpha=-1.5$.
Initial conditions are: $M(0)=11\,$TeV$/c^2$ and $p(0)=0.01\,$TeV$/c$.
N/A means black hole mass does not grow.
\label{L_dependence_p001}}
\end{table*}
\begin{table*}
\centering
\begin{tabular}{|c|c|c|c|c|c|c|c|c|c|}
\hline
$L$ ($\mu$m)&$\alpha$& $\mc$  (TeV$/c^2$)
&$dM/dt|_{t=0}$& $M$  (kg)& $\rem$ (m) & $\rh$ (m)
\\
\hline
44&-1.44& $2\cdot 10^{53}$& $2.8\cdot 10^{-2}$ & $4.0\cdot 10^{-13}$
&$1.3\cdot 10^{-14}$ &$9.0\cdot 10^{-25}$
\\
\hline
5.0& -1.50&$4\cdot 10^{53}$& $1.1\cdot 10^{-2}$ & $1.9\cdot 10^{-13}$
&$6.5\cdot 10^{-15}$ &$2.0\cdot 10^{-25}$
\\
\hline
$1.0\cdot 10^{-1}$ &1.59&$1\cdot 10^{51}$& $1.3\cdot 10^{-3}$
& $5.3\cdot 10^{-14}$&$1.7\cdot 10^{-15}$ &$1.8\cdot 10^{-26}$
\\
\hline
$1.0\cdot 10^{-2}$ &-1.67& $1\cdot 10^{54}$& $4.5\cdot 10^{-4}$
& $2.4\cdot 10^{-14}$ &$8.1\cdot 10^{-16}$ & $2.7\cdot 10^{-27}$
\\
\hline
\end{tabular}
\caption{Time evolution of black hole mass with critical mass $\mc$ near upper bound
for time equal to approximative age of the Universe and $\beta=1.1$.
Initial conditions are: $M(0)=11\,$TeV$/c^2$ and $p(0)=0.0001\,$TeV$/c$.
\label{mcmax0001}}
\end{table*}
We then proceed to study the region $1<\beta\lesssim 1.25$,
in which the mass of the black hole can grow.
\par
\begin{figure}[b]
\centering
\raisebox{1.6in}{$M$}
\epsfxsize=3.0in
\epsfbox{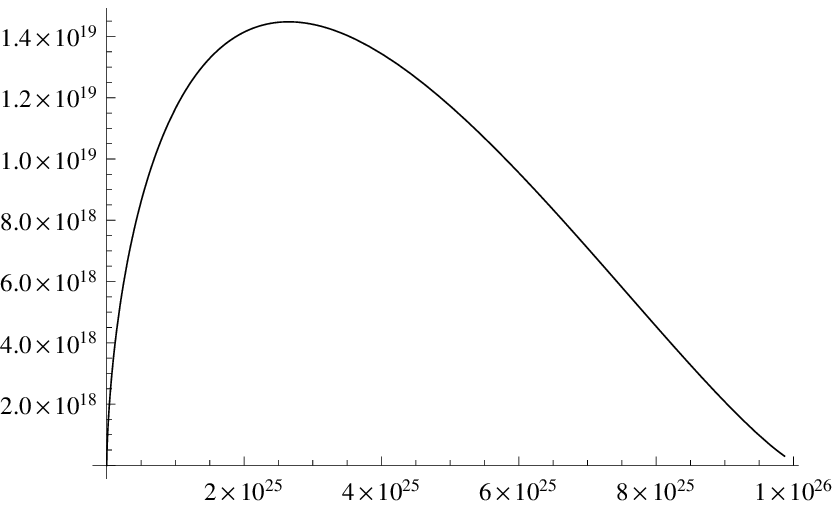}
$t$
\\
\raisebox{1.5in}{$p$}
\epsfxsize=3.0in
\epsfbox{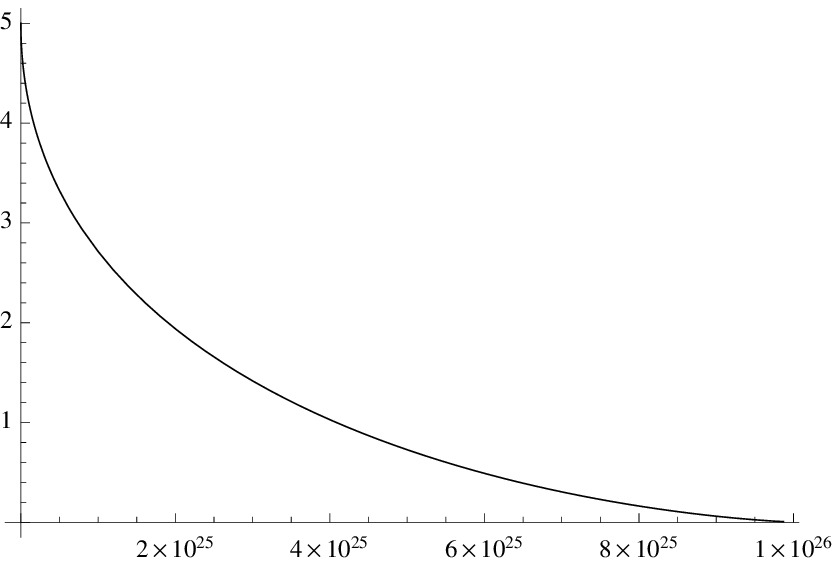}
$t$
\caption{Mass (in TeV$/c^2$) and momentum (in TeV$/c$)
for $L=44\,\mu$m, $\beta=1.25$, $\alpha=-1.8$, $M(0)=10\,$TeV$/c^2$
and $p(0)=5\,$TeV$/c$.}
\label{slow}
\end{figure}

The evolution of the black hole mass as a function of the
initial mass and momentum is shown in Table~\ref{p_0},
for a constant thickness of the brane $L=44\,\mu$m, $\beta=1.25$,
and $\alpha=-1.8$.
The critical mass~\eqref{mc} for this choice of parameters is
$M_c=10^{44}\,$TeV$/c^2$,
which is within the allowed range.
A typical example is also plotted in Fig.~\ref{slow}.
\par
One can see that the maximum value of the black hole mass decreases
as the initial momentum decreases.
This could already be inferred from Eq.~\eqref{acc}.
In fact, the accretion rate is proportional to the black hole velocity and,
for lower velocities, the accretion rate decreases and the evaporation rate
becomes more and more dominant.
We stress that the maximum mass was calculated assuming that the black hole
would travel through a medium with a density equal to the average density of the Earth
all the distance from the point of creation to the point of maximum mass.
\par
For black holes created on Earth, the maximum value of the mass $M_{\rm E}$
would indeed be much smaller, since after crossing the Earth, the density drops to zero
and so does the accretion rate.
From Table~\ref{p_0}, the actual value of the mass when the black hole leaves
the Earth is on average fifteen orders of magnitude smaller than the potential
maximum mass and its capture radius $\rem$ too small to start Bondi accretion.
Another point that needs to be remarked is that, for most values of the initial
momentum, the black hole crosses the Earth with a residual velocity $v_{\rm E}$
larger than the escape velocity ($\simeq 11\,$km$/$sec) and can in fact leave
our planet.
The only case in which the velocity is smaller than the escape velocity occurs
for $p(0)=0.01\,$TeV$/c$, but in this case the maximum mass is just
of the order of $10^{-9}\,$kg.
In all cases, the capture radius $\rem$ remains much larger than the gravitational
radius $\rh$, ensuring the Newtonian approximation for the former holds.
\par
Given the dependence of the results on the initial momentum,
we next analyze separately the regimes with large or small initial momentum.
\subsubsection{Large initial momentum}
\label{large_p0}
The data in Table~\ref{L_dependence} shows the dependence of the maximum
black hole mass on the thickness of the brane for $p(0)=5\,$TeV$/c$,
$M(0)=10\,$TeV$/c^2$, $\beta=1.1$ and $\alpha=-1.5$.
Note that the constraint in Eq.~\eqref{alpha_sun} excludes a thickness
$L\simeq 44\,\mu$m.
The maximum attainable mass, if the black hole would travel through matter
with a constant density equal to the Earth's, is directly proportional to $L$
in this case and tops at $L\approx 5\,\mu$m.
The value of the black hole mass when leaving the Earth in this case is of the
order of $10^{-21}\,$kg.
Below $L=0.01\,\mu$m the black hole decays instantaneously again.
\par
Table~\ref{mcmax} was obtained with the parameters adjusted so as to keep
the critical mass $\mc$ near the maximum allowed value
$M_\odot\simeq 10^{54}\,$TeV.
Again the initial value for the momentum was set to $p(0)=5\,$TeV$/c$
and the initial mass of the black hole to $M(0)=10\,$TeV$/c^2$.
One can then observe that the potential maximum mass $\mmax$
is again very large
but the time taken to reach it is at least $10^{62}\,$sec, much larger
than the estimated age of the Universe ($\simeq 10^{18}\,$sec).
The data in the table also shows that the black hole would cross our planet
in seconds with a final velocity $v_{\rm E}$ much larger than the Earth's
escape velocity.
The mass $M_{\rm E}$ of the black hole at that time is on the order of $10^{-22}\,$kg,
which is again very small, and its capture radius $R_{\rm E}$
well below the scale of Bondi accretion.
\subsubsection{Small initial momentum}
\label{small_p0}
Similar dependencies were studied for the case of small initial momentum.
In Table~\ref{L_dependence_p001}, the initial momentum of the black holes was set to
$p(0)=0.01\,$TeV$/c$ and the initial mass $M(0)=11\,$TeV$/c^2$.
The maximum attainable black hole mass is smaller than in the case with a larger initial
momentum, and the black hole leaves the Earth with velocity larger than the escape velocity
from the gravitational field of the Earth.
\par
If one studies the black hole velocity as a function of the initial momentum with the
critical mass $\mc$ kept near its maximum allowed value, one finds that the highest value
of the initial momentum for which the velocity of the black holes after passing through
the Earth is smaller than the escape velocity is on the order of $100\,$MeV/c.
Considering the data in Table~\ref{p_0}, in order the maximize the maximum mass
that the black hole can reach, its initial momentum needs to be near the highest
possible value.
Table~\ref{mcmax0001} was obtained setting the parameters so that the critical mass
is near the maximum allowed value, with $p(0)=10^{-4}\,$TeV$/c$ and
$M(0)=11\,$TeV$/c^2$ respectively.
The black hole in this case has the highest initial momentum which is small enough
to be trapped inside the Earth.
Due to the extremely large black hole lifetimes encountered in this case,
we studied the evolution for a duration of the order of magnitude of the present age
of the Universe and found a final mass of the order of $10^{-14}\,$kg.
As in all the previous cases, the capture radius remains much smaller than
the atomic size and the black hole does not start Bondi accretion.
Also the gravitational radius is orders of magnitude smaller than the electromagnetic radius,
which means that the Newtonian approximation used to derive the accretion rate holds.
\section{Conclusions}
\label{conc}
We studied the evolution in time of microscopic black holes which could be
produced at the LHC, based on the model presented in Refs.~\cite{CH,bhEarth1}
and the description of brane-world black holes given in Ref.~\cite{dadhich}.
In particular, we extended the treatment of Ref.~\cite{bhEarth1} by considering
a general form of the tidal term containing two parameters, $\alpha$ and $\beta$,
whose ranges were first analyzed in Section~\ref{metrics}.
The parameter $\beta$ was allowed to take any positive value.
Conditions for the black holes to be tidal and phenomenologically acceptable
were then used to determine the range for $\alpha$ at fixed $\beta$.
Subsequently, the time evolutions of the black hole mass and momentum were
obtained numerically with the two parameters $\alpha$ and $\beta$ in the
allowed ranges by solving the equations which govern the luminosity and
accretion rates, in Sections~\ref{evaporation}-\ref{evol}.
\par
First, we found that tidal black holes would evaporate (almost) instantly,
except for $1<\beta\lesssim1.25$.
(The particular case with $\beta=1$ was studied in Ref.~\cite{bhEarth1}.)
Two distinct regimes were then taken into consideration inside this range:
large initial momentum, and small initial momentum.
Numerical data for the regime with large initial momentum are presented in
Tables~\ref{L_dependence} and~\ref{mcmax}, and show that the black holes
with a large value of the initial momentum would cross the Earth
in a matter of seconds and come out with velocities much larger than the Earth's
escape velocity.
Their mass, after crossing the Earth, is of the order of $10^{-22}\,$kg, after which
accretion turns off, and the black holes just evaporate.
If the black holes are created with a small initial momentum, it is possible that
they are trapped inside the Earth.
However, Table~\ref{p_0} shows that the maximum mass decreases for
decreasing initial momentum.
Therefore, the absolute maximum mass is reached for the maximum initial
momentum which is still small enough to allow for trapping.
Tables~\ref{L_dependence_p001} and~\ref{mcmax0001} then show that,
for black holes trapped inside the Earth, after a time comparable with the
age of the present Universe, the mass is on the order of $10^{-14}\,$kg,
which is still negligibly small.
\par
Our overall conclusion is therefore that the tidal-charged black holes are a viable
model of micro-black holes which might be produced at the LHC.
The model predicts that such black holes cannot grow to catastrophic size,
but might live long enough to escape the detectors and result in significant
amounts of missing energy.
\end{document}